# Healthcare Security Breaches in the United States: Insights and their Socio-Technical Implications

**Megha Moncy**
Department of BioHealth Informatics
Luddy School of Informatics, Computing and Engineering,
Indiana University Indianapolis, IN, USA.

**Sadia Afreen**
Department of BioHealth Informatics
Luddy School of Informatics, Computing and Engineering,
Indiana University Indianapolis, IN, USA.

**Saptarshi Purkayastha**
Department of BioHealth Informatics
Luddy School of Informatics, Computing and Engineering,
Indiana University Indianapolis, IN, USA.

## ABSTRACT

This research examines the pivotal role of human behavior in the realm of healthcare data management, situated at the confluence of technological advancements and human conduct. An in-depth analysis spanning of security breaches in the United States from 2009 to the present elucidates the dominance of human-induced security breaches. While technological weak points are certainly a concern, our study highlights that a significant proportion of breaches are precipitated by human errors and practices, thus pinpointing a conspicuous deficiency in training, awareness, and organizational architecture. In spite of stringent federal mandates, such as the Health Insurance Portability and Accountability Act (HIPAA) and the Health Information Technology for Economic and Clinical Health (HITECH) Act, breaches persist, emphasizing the indispensable role of human factors within this domain. Such oversights not only jeopardize patient data confidentiality but also undermine the foundational trust inherent in the healthcare infrastructure. By probing the socio-technical facets of healthcare security infringements, this article advocates for an integrated, dynamic, and holistic approach to healthcare data security. The findings underscore the imperative of augmenting technological defenses while concurrently elevating human conduct and institutional ethos, thereby cultivating a robust and impervious healthcare data management environment.

# INTRODUCTION

In the burgeoning digital world, healthcare data is at a critical juncture between innovation and susceptibility. Ensuring secure and safeguarded data handling has emerged as paramount, given the vast repositories of sensitive personal information inherent in healthcare (Gostin et al. 2009). The rapid adoption of EHR systems in the US in the last decade (from less than 9% physicians using an EHR in 2009 to over 98% in 2022) due to legislative changes (Adler-Milstein et al. 2014; Holmgren et al. 2022), has produced large electronic databases, but human systems usually change much slowly (Neff 2017). This mismatch is the cause of a number of socio-technical challenges, including in information security. Nonetheless, the increasing incidence of data breaches, stemming from theft or system failures, emphasizes an urgent need to bolster security measures and devise resilient data protection frameworks across systems (Yaacoub et al. 2020). The 2023 ForgeRock report noted a substantial 136% surge in data breaches from the previous year. Such breaches inflict a significant financial burden on affected entities and result in marked deterioration in their market standing (Ko and Dorantes 2006). The rapid growth of data breach occurrences remains alarming, with a disproportionate concentration in the healthcare sector (Seh et al. 2020). This financial burden due to data breaches, makes a health system like the United States, already the most expensive with poor health outcomes, even more expensive. The aggregate effects of security lapses, inadvertent or deliberate data disclosures, and other organizational vulnerabilities contribute to the escalating trend (McLeod and Dolezel 2018). However, a confluence of socio-technical determinants, such as technical inefficiencies, legal challenges in caregiver data handling, insufficient staff training, and management development issues, offers a comprehensive lens to discern the multifaceted nature of data breaches (Kamoun and Nicho 2018).

## Our Contribution:

In our research, we examined healthcare data breach disclosures in the United States in the last decade, with a special emphasis on understanding a reciprocal relationship between technology and people (socio-technical) dynamics and the escalating cyber threats. Prior research using quantitative analysis of the breaches has shown human factors in health data breaches (Yeo and Banfield 2022), and individual case studies have identified socio-technical causes of a single data breach (Lee and Choi 2021). However, very little is known at the health system level of study (higher than organizational but lower than societal), on socio-technical explanations and theorization of rapidly growing health data breaches. We notably shed light using qualitative data analysis of the breach notifications. In addition to identifying vulnerabilities, we aim to highlight key factors that strengthen security measures, emphasizing the importance of qualitative insights in developing resilient cybersecurity solutions for healthcare data (Phillips et al. 2020; Purkayastha et al. 2020). Specifically, the research seeks to answer the following questions:

1. What are the key socio-technical implications of healthcare security breaches, and their impact on confidentiality and integrity of patient data?
2. To what extent do socio-technical factors contribute to the escalating trend of healthcare data breaches, and how can a holistic understanding of these factors lead to more impactful cybersecurity solutions at the health system level?
3. How does the geographical distribution of healthcare data breaches within the United States reflect variations in cybersecurity breaches?
4. What role do human behaviors and organizational cultures play in mitigating or exacerbating healthcare security breaches?

# METHODOLOGY

We employed a mixed-methods approach in an explanatory sequential design (Fetters et al. 2013). This methodology was structured in two primary phases: a quantitative analysis of breach notification reports and a qualitative discourse analysis of text descriptions. The source of the data was breach notification reports from the U.S. Department of Health and Human Services since 2009.

**Data collection:**

Our principal data source is the Breach Notification Portal [1], frequently termed the "Wall of Shame.", is an authoritative record of healthcare data breach incidents in the United States, with the 13402(e)(4) of the HITECH Act requires the Secretary to post a list of breaches of unsecured protected health information affecting 500 or more individuals. In the notification portal, each row represents a unique breach event with Breach identification number, Date of breach, Entity involved, Type of breach (e.g., hacking, theft), Number of records affected and Textual description of the breach event.

**Dataset Composition:**

Using the portal archive, we extracted the full dataset spanning from 2009 to the present, comprising 4751 breach notifications. However, this does not include full descriptive reports of cases that are pending investigations, which must be completed, and their results posted within 24 months of the breach notification. This breach description served as a principal analysis for our study. This expansive and detailed dataset not only enables an evaluation of the current state of healthcare data breaches but also facilitates an exploration of their evolution over time.

## Data Analysis:

---

[1] https://ocrportal.hhs.gov/ocr/breach/breach_report.jsf

**Quantitative data analysis:**

Before any analytical operations, the dataset underwent a thorough cleaning process to handle missing values, remove duplicates, and standardize entries for consistency. Descriptive statistics were computed to understand the distribution, frequency, and patterns of breaches across different years, entities, and types of breaches with findings similar to Walden et al. (2020). The number of records affected in each breach was also analyzed to gauge the magnitude and impact of these breaches over the decade.

**Qualitative data analysis:**

Given the extensive size of the dataset, a stratified random sampling method was used to select textual descriptions. This ensured a representative sample across different types of breaches, years, and entities. Discourse analysis was employed to uncover the underlying socio-technical narratives present in the text descriptions of breaches. This analysis aimed to identify recurrent themes and patterns in the language used to describe breaches, understand the socio-technical contexts in which these breaches occurred and extract insights on human and organizational behaviors leading to or exacerbating the breaches. The text descriptions were first coded using open coding to identify emergent themes. These initial codes were then refined and grouped into broader categories. The inter-coder reliability was maintained by having two independent researchers code a subset of the data and then reconcile any discrepancies through discussion. Post-analysis, the findings from both the quantitative and qualitative analyses were integrated to provide a holistic understanding of healthcare security breaches and their socio-technical implications. This integrative approach allowed for a nuanced interpretation, combining the statistical patterns evident in the breach reports with the richer context and narratives captured through discourse analysis. We took a 5-step process for the qualitative data analysis:

1. Familiarization: Multiple researchers read through a subset of the descriptions to become familiar with the content.
2. Generating Initial Codes: We identified recurring themes or patterns in the descriptions.
3. Searching for Themes: We grouped related codes into broader themes.
4. Reviewing Themes: We refined and consolidated themes.
5. Defining and Naming Themes: We clearly defined what each theme encompasses.

Several recurring themes and patterns were observed by an individual reviewer and then clubbed by the reviewers. *Nature of Breach*: Descriptions often specify the type of breach, such as email phishing attacks, uploading to public-facing websites, ransomware attacks, etc. *Entities Involved*: There are mentions of covered entities (CE) and sometimes business associates (BA), indicating the entities involved in the breach. *Protected Health Information (PHI) Compromised*: Each description provides details about the types of information compromised, such as names, addresses, birthdates, social security numbers, clinical information, and more. *Response and Mitigation*: Descriptions include steps taken by entities upon discovering the breach, such as disabling email accounts, removing files from public websites, retraining staff, etc. *Notification*: There's a mention of entities providing breach notification to various stakeholders, including HHS, affected individuals, and sometimes the media. *Impact and Outcome*: Some descriptions provide insights into the consequences of the breach, such as the provision of free credit monitoring services, termination of involved workforce members, or pursuit of criminal charges.

Some examples of initial codes in each category included:

1. Nature of Breach: phishing, public-facing website, ransomware, downloaded data, employee error, business associate error.

2. Entities Involved: covered entity, business associate, workforce member.

3. PHI Compromised: names, addresses, birthdates, social security numbers, clinical information.

4. Response and Mitigation: disabled email accounts, removed from website, retrained staff, revised technical deployment, revised practice, new safeguards.

5. Notification: notification to HHS, affected individuals, media, free credit monitoring.

6. Impact and Outcome: staff termination, criminal charges, anomalies, conflicts of interest, confidentiality of PHI.

At an aggregate, the Nature of Breach codes were applied 4259 times, Entities Involved: 5452 times, PHI Compromised: 6819 times, Response and Mitigation: 550 times, Notification: 8121 times, Impact and Outcome: 197 times. The high frequency of codes related to "PHI Compromised" and "Notification" suggests that these areas are predominant in the web descriptions. The themes related to "Response and Mitigation" and "Impact and Outcome" have comparatively fewer occurrences, indicating that while these areas are addressed in some descriptions, they might not be as consistently detailed as other themes. The refined themes provide a structured framework to understand the myriad facets of healthcare security breaches. To deepen our understanding and gain more insights, we can further analyze a subset of descriptions for each theme, identifying nuances, patterns, and any additional sub-themes that might emerge. There have been limited qualitative analysis of healthcare data breaches at the health system level (Pool et al. 2024; Zou et al. 2019), and the most common qualitative approach has been content analysis with news reports (Hammouchi et al. 2019) or forensic analysis during specific breach investigations.

# FINDINGS

**Quantitative Findings:**

Our analysis demonstrated a substantial increase in the number of reported healthcare security breaches over the past decade. This trend underscores the growing vulnerabilities and challenges in healthcare data management in the digital age. <u>Entities Involved</u>: Covered entities (CEs) were frequently associated with breaches, but there was also a significant involvement of business associates (BAs), highlighting the interconnectedness of the healthcare data ecosystem and the need for rigorous data handling and sharing protocols. <u>Nature of Breaches</u>: Phishing attacks, unauthorized access, and hacking incidents were predominant. The recurring nature of these breaches points towards consistent vulnerabilities in healthcare data systems and human behavior. <u>PHI Compromised</u>: The most commonly compromised data included basic identifiers (names, addresses, birthdates), followed by more sensitive identifiers like social security numbers and medical records. This shows the depth of information vulnerable in such breaches, making the implications even more severe.

***Data breach frequency and impact on the individual***: We observed a significant disparity in the frequency and impact of data breaches across distinct healthcare entities. The first pie chart indicates healthcare providers as the most frequently breached entity and therefore emphasizes a vulnerability within this sector. Whereas business associates and health plans depict a notable but comparatively lower breach occurrence.

Furthermore, we gauge the severity of these breaches by analyzing the proportion of affected individuals within each entity type. Healthcare providers again stand out, showcasing a critical need for enhanced security measures. Following which, health plans closely trail, reflecting a substantial impact on insurance and coverage data. Business associates, and healthcare clearinghouses though less impacted with a minor frequency, still have an effect on individuals, indicating a need for vigilance across the board.

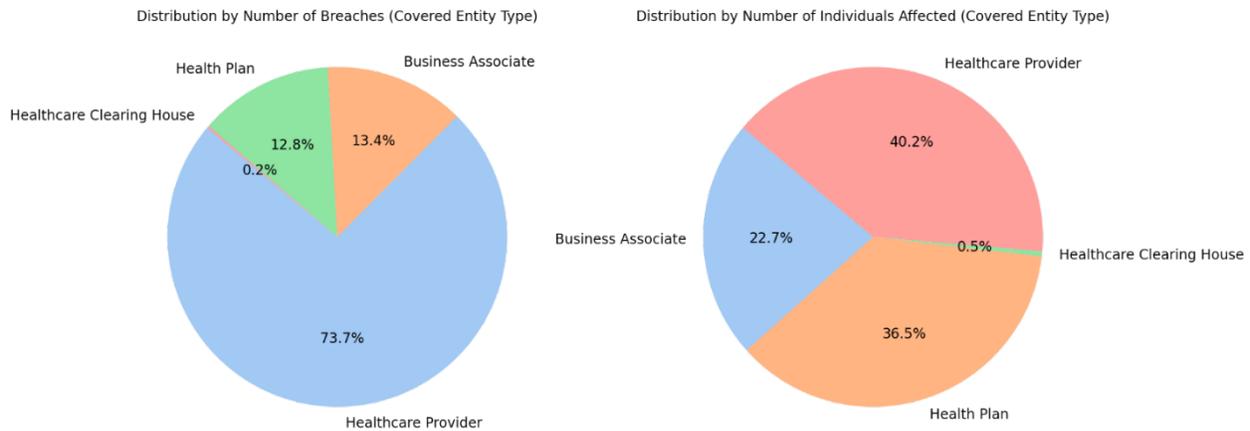

**Figure 2.** Analysis of number of breaches vs number of individuals affected.

***Data breach distribution by State***: We further observed a notable disparity in the distribution of data breaches across different states in the United States. California (CA) and Texas (TX) appear as the most heavily impacted states, indicating a concerning vulnerability to data breaches. On the contrary, states like Idaho (ID), South Dakota (SD), and North Dakota (ND) display significantly lower breach occurrences.

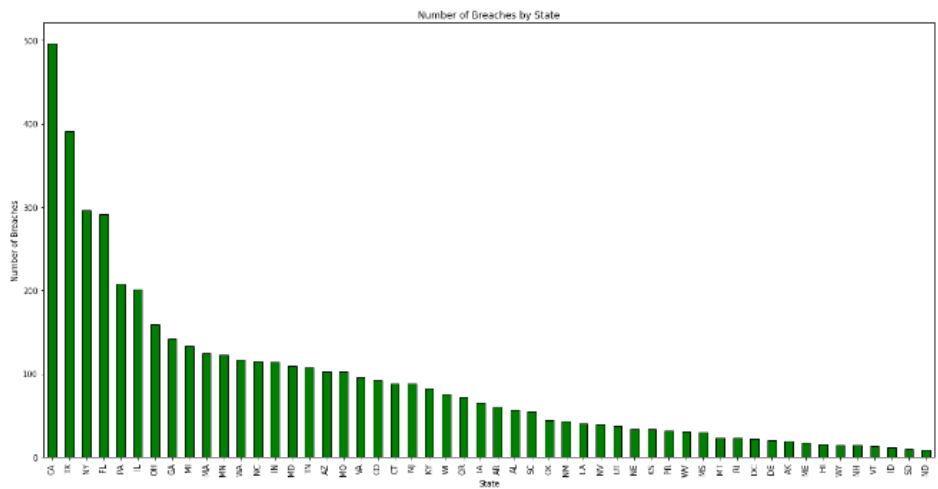

**Figure 3.** Distribution of breach frequency by U.S. State

*Impact of data breach types on individuals:* With the help of box plot, we visually compared different types of data breaches based on the number of individuals affected. The Hacking/IT incidents showcased a wide range, indicating that some breaches impacted a lot more individuals compared to others. On the other hand, breaches due to unauthorized access or disclosure had a smaller yet more consistent impact. Theft-related breaches were similar in this aspect too. Improper disposal breaches had a broader impact range, meaning their consequences varied more. Lastly, breaches categorized as loss had a more consistent impact range. This comparison helped us understand the varying impact of different types of breaches on individuals.

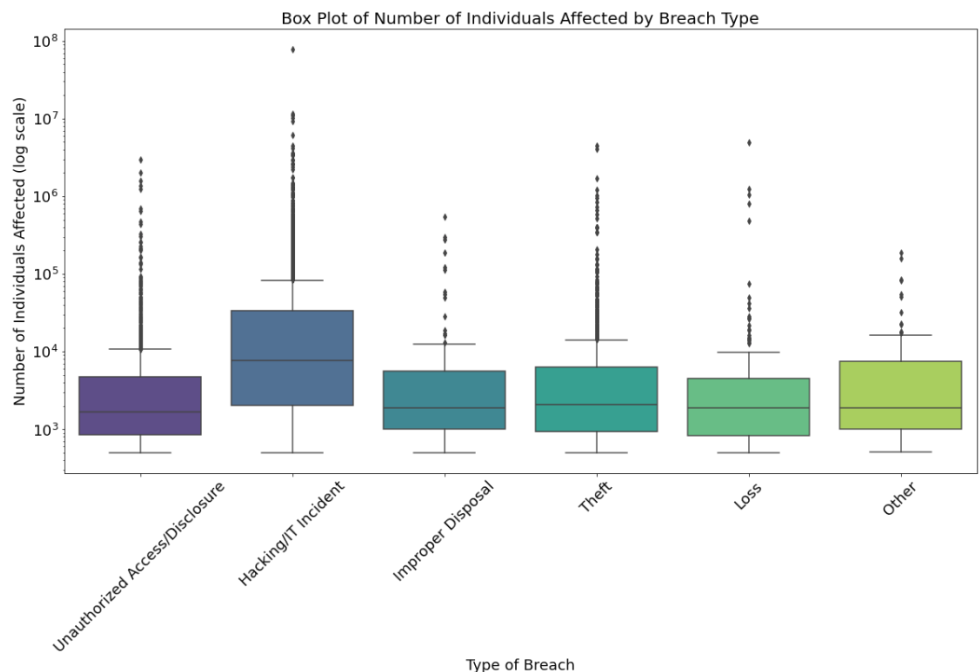

**Figure 4.** Number of individuals affected by breach type.

*Data breaches over the years:* In our analysis, we utilized a chart to present the yearly and monthly growth trendlines concerning the number of breaches. The monthly trendline displayed a turbulent path, indicating frequent fluctuations and an overall consistently inclining pattern in breach occurrences. The breach is seen to have maximized over the years, up until 2021, after which further years data is yet under investigation. These trends provide insights into how data breaches are growing rapidly, emphasizing the need for heightened security measures and vigilant monitoring to mitigate risks effectively.

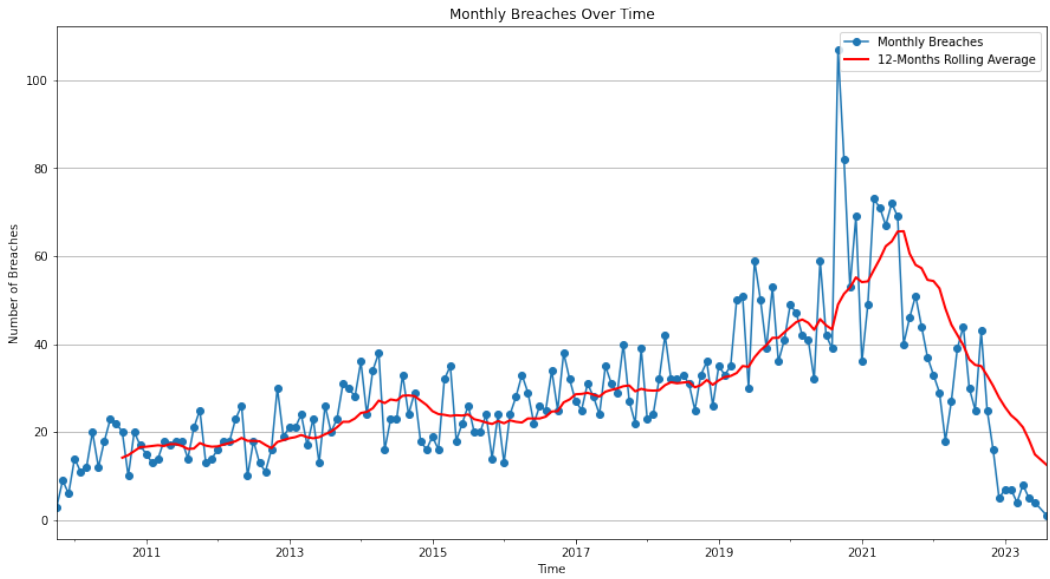

**Figure 5.** Number of individuals affected by healthcare breach.

**Qualitative Findings:**

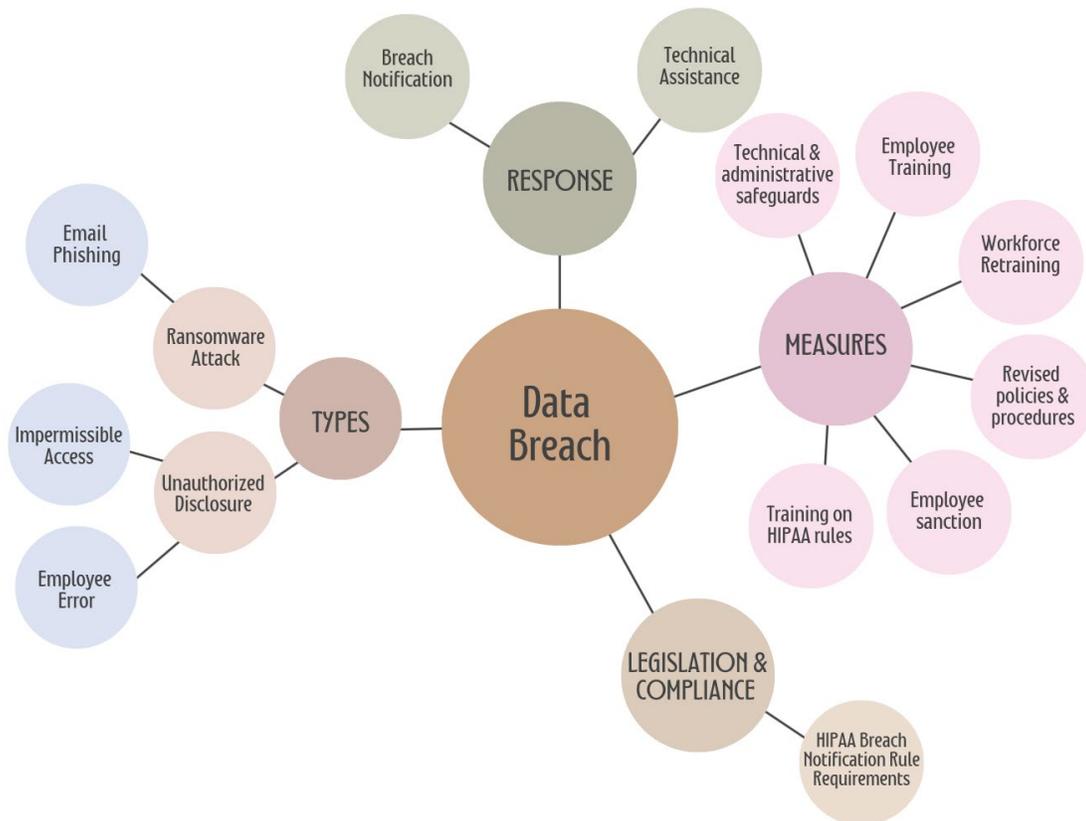

**Figure 6**. Thematic exploration of data breaches

***Categorizing Human Errors:*** <u>Physical Theft</u>: Instances where physical devices like desktop computers, laptop computers, and backup drives were stolen. <u>Mailing Errors</u>: Breaches resulting from mailing sensitive information to incorrect recipients. <u>Unauthorized Access</u>: Breaches where former workforce members or unauthorized individuals accessed medical record storage facilities. <u>Cyber Attacks</u>: Instances of ransomware attacks that used phishing emails to install backdoors that resulted in ransomware installation. <u>Business Associate Issues</u>: Breaches attributed to issues with a business associate, such as fraudulent provider accounts.

***Technical safeguards***: revise deployment practices for replacing outdated systems that result in vulnerable access to PHI (Handley et al. 2023). As the healthcare landscape increasingly embraces advanced medical technologies, a looming threat arises from outdated systems. Often incompatible with contemporary security standards, these systems create an expansive opening for cyber-attackers, putting the integrity of protected health information (PHI) at grave risk (Tully et al. 2020). To counter this vulnerability, healthcare institutions must prioritize the adoption of enhanced technical safeguards. This involves transitioning to up-to-date platforms fortified with robust administrative protections. Collaborating with tech vendors who specialize in healthcare IT ensures access to cutting-edge protective measures, regular software refreshes, and state-of-the-art encryption mechanisms, ensuring PHI remains shielded from unauthorized incursions.

***Human element***: Administrative safeguards and workforce dynamics sometimes result in inadequate security management and unauthorized access (Baran and Woznyj 2020). This fundamental issue lies in the inadequacy of security management processes. The absence of comprehensive procedures leaves healthcare organizations vulnerable to breaches and unauthorized access to ePHI. The laxity in addressing prevention, detection, and correction of security violations further exacerbates the problem.

***Enhanced security management processes and assigning a Privacy Officer***: To bolster security defenses, healthcare institutions must prioritize administrative safeguards by establishing robust security management processes (Malloy 2003). These procedures must lay a strong foundation for HIPAA compliance, emphasizing risk analysis to pinpoint vulnerabilities and assess the probability of breaches. Formulating internal policies to impose stringent penalties for non-compliance is essential to reinforcing the security infrastructure. Also, organizations should ensure the appointment of a privacy officer, as mandated by HIPAA. This officer will helm the development and regular update of security policies and procedures, thereby ensuring adherence to the highest standards of data protection.

***Neglect of security training and employee error***: The neglect of employee training in HIPAA security standards contributes to errors, creating loopholes for potential cyber-attacks and unauthorized disclosures (Hatzivasilis et al. 2020). A common approach taken in breach notification is mandatory HIPAA security training, which is anyway done commonly. However, since it's usually considered a "chore" many health workers rarely pay practical attention to these trainings (Yeng et al. 2022). Retraining of staff, especially in areas like email security, is a common measure taken. A robust security awareness training program for all staff members will impart essential knowledge about the significance of security, protection mechanisms against malicious software, and best practices for password management, significantly reducing the chances of employee errors leading to breaches.

## DISCUSSION AND CONCLUSION

Our findings emphasize the intricate intersection of technological vulnerabilities and human behavior in the context of healthcare security breaches.

***Human behavior as a vulnerability***: The prominence of breaches stemming from phishing attacks and unauthorized access, as revealed in our findings, underscores the human component as a significant vulnerability in the healthcare data ecosystem. This aligns with the observations of Nifakos et al. (2021), who argued that while technology continues to advance, human behavior remains a constant vulnerability, often exploited in cyberattacks. The ease with which employees can be deceived by well-crafted phishing emails or social engineering tactics highlights the need for continuous and adaptive training programs tailored to the evolving threat landscape.

***Socio-Technical thinking in healthcare***: The intertwined nature of technology and human behavior in healthcare settings can be better understood through the lens of socio-technical systems. As Baxter and Sommerville (2011) noted, socio-technical systems consider both the technical aspects (software, hardware, network configurations) and the social aspects (work routines, organizational structures, user behaviors) of a system. Our findings, which show breaches resulting from both technological flaws and human errors, emphasize the need to view healthcare data systems as socio-technical entities. Addressing vulnerabilities thus requires solutions that encompass both the technical and the human elements.

***Organizational culture and security awareness***: The instances of breaches involving business associates, as highlighted in our research, indicate gaps in organizational culture and awareness across the healthcare data chain (Purkayastha et al. 2015). This observation resonates with the findings of Akter et al. (2022), who argued that an organization's culture plays a pivotal role in its cybersecurity posture. Fostering a culture that prioritizes data security, not just within a single entity but across all partners and stakeholders, is crucial.

***Holistic approaches to data security***: The discourse analysis from our research, illustrating the interconnectedness of various breach themes, echoes the sentiment of Ulrich et al. (2021), who advocated for holistic approaches to cybersecurity. Addressing isolated vulnerabilities might offer temporary respite, but a comprehensive strategy that considers both technological and socio-technical aspects ensure a more resilient system.

***The role of training and continuous learning***: Our findings underscore a significant gap in training and awareness. This gap is further emphasized by Rajamäki et al. (2018), who found that continuous learning and periodic training sessions for employees significantly reduce the risk of breaches stemming from human errors. As threat actors continually refine their tactics, healthcare entities must equally prioritize ongoing education to stay a step ahead.

***Limitations, biases and ethical considerations of our study***: The study exclusively used publicly available data, ensuring that there were no ethical concerns related to data privacy or informed consent. Nonetheless, all data were treated with the utmost confidentiality, and any potentially identifiable information was anonymized during the analysis. While the mixed-methods approach provides a comprehensive view of healthcare security breaches, it is essential to note the inherent limitations. The reliance on reported breaches might introduce bias, as not all breaches might have been reported or captured. The discourse analysis, being interpretative, is subject to the researchers' perspectives, even though measures like inter-coder reliability were employed to minimize subjectivity.

Combining quantitative and qualitative analyses, was designed to offer in-depth insights into the healthcare security breaches in the U.S. and their socio-technical implications.